# Stoichiometric oxygen content in $Na_xCoO_2$

L. Viciu,[1] Q. Huang[2] and R. J. Cava[1]

[1] Department of Chemistry, Princeton University, Princeton NJ 08540

[2] NIST Center for Neutron Research, NIST, Gaithersburg, MD 20899

## Abstract

The crystal structure and oxygen stoichiometry in two-layer $Na_{0.74}CoO_2$ and $Na_{0.38}CoO_2$ at room temperature are analyzed by powder neutron diffraction. Two sets of diffraction data for each sample, taken at different incident neutron wavelengths, $\lambda$=1.1968Å and $\lambda$=1.5403Å, are analyzed simultaneously by the Rietveld method, allowing for the independent refinement of all structural parameters. The fractional oxygen site occupancies are found to be 1.01(1) for $Na_{0.74}CoO_2$ and 0.99(2) for $Na_{0.38}CoO_2$ respectively. These results indicate that the oxygen content of these phases is stoichiometric to a precision of 1 to 2%, and therefore the formal cobalt oxidation state is determined solely by the sodium content. The analysis also reveals that both types of sodium ions in the structure are in off-center distorted trigonal prismatic geometry.

## Introduction

The structural variety and diversity of properties in the $Na_xCoO_2$ layered sodium cobalt oxides have made them the subject of continuing research for both chemists and physicists. Their magnetic and electronic properties are nominally tuned by changing the number of charge carriers in the $CoO_2$ layers through sodium intercalation or deintercalation.[1, 2, 3] The recent discovery of superconductivity in $Na_{0.3}CoO_2$ intercalated with water has further suggested that unconventional superconductivity and novel quantum states arising from strong electron correlations may occur in this family.[4] The origin of superconductivity in this material is not yet fully understood.

The electronically active structural components of $Na_xCoO_2$ are sheets of edge-sharing $CoO_6$ octahedra. The Co layers have a triangular lattice, and close packed triangular layers of oxygen are found both above and below the Co plane. Each oxygen in the close packed oxygen layer is bonded to three Co in the neighboring Co layer and 1-3 Na in the interleaving Na plane, depending on the Na stoichiometry. The $Na_xCoO_2$ electronic phase diagram shows that a paramagnetic metal is found for $0.5<x<0.75$, an insulator is found for $x=0.5$, and a normal metal is found for $x<0.5$.[5] Critical to any interpretation of the properties of the system is knowledge of the formal charge in the $CoO_2$ layers. This has generally been taken to be determined by measurement of the Na concentration and the requirement of charge neutrality to yield formal Co oxidation states given by $Na_xCo^{+(4-x)}O_2$. In the range of interesting properties, $0.3<x<1.0$, the oxidation state of Co is then between +3.7 and +3.0. Early in the studies of the superconducting oxyhydrate, however, and also in studies of the simpler non-hydrated $Na_xCoO_2$ phases, the use of the Na content as a measure of the Co oxidation state was called into question.

These questions have been raised primarily based on the interpretation of chemical titration data, which, in the ideal case, measures the oxidizing power of solid particles suspended in a liquid containing an indicating reagent. Precision application of this method requires excellent technique, rigorously single-phase powder samples with fine particle size, and unambiguous knowledge of the chemistry of the reactions. Such methods have yielded lower Co oxidation states in these materials than are expected from the nominal compositions. Thus, titration-based analyses have suggested that for the superconducting phase, $Na_xCoO_2 \times 1.3H_2O$, the oxidation state of cobalt is between +3.3 and +3.5, which is substantially lower than what is expected from the sodium content.[6,7] Time dependent neutron diffraction analysis of the superconducting sample indicated that the Co oxidation state changes in time, and the maximum Tc occurs when Co is in ~+3.5 oxidation state.[8] These discrepancies have been attributed to the presence of $H_3O^+$ or $H^+$, whose presence has not been detected directly, or to oxygen vacancies. Titration analyses have been performed on unhydrated $Na_xCoO_2$, with the conclusion that no further oxidation of Co is possible beyond Co +3.5.[9] In the simple ternary compound, then, this means that oxygen would have to be removed from the close-packed layer under oxidizing conditions, i.e. that for x<0.5 the formula of the compound is $Na_xCo^{+3.5}O_{1.75+x/2}$ and the Co oxidation state is constant at +3.5. In one instance, thermogravimetric analysis of a high x, $Na_{0.75}CoO_2$ sample, was interpreted to indicate an oxygen deficiency of ~0.08, or a Co oxidation state of +3.09 when one of +3.25 is expected from the Na content.[10]

Here we present the results of careful neutron diffraction measurements on $Na_xCoO_2$ in both the high x and low x regions, designed specifically to be sensitive to the presence or absence of oxygen vacancies. The quality of the data and the model fits are excellent. No oxygen vacancies are found. Combined with the results of recent high resolution neutron diffraction measurements on $Na_{0.5}CoO_2$,[11] it can be concluded that the Na content is a good measure of the formal oxidation state in $Na_xCoO_2$ and, therefore, a measure of the charge in the $CoO_2$ layer in the range of interesting compositions.

**Experimental**

$Na_{0.74}CoO_2$ was prepared by solid-state reaction. Stoichiometric quantities of $Co_3O_4$ (Alfa, 99.99%) and $Na_2CO_3$ (Alfa, 99.9985%) were ground together and annealed at 800°C for 16 hours under flowing $O_2$. 5% excess of $Na_2CO_3$ was added to balance the Na lost due to volatilization. $Na_{0.38}CoO_2$ was prepared by stirring 1 g of $Na_{0.74}CoO_2$ with 40X $Br_2$ dissolved in acetonitrile. After five days of stirring at room temperature, the product was washed with acetonitrile and stored under dry conditions to prevent water intercalation.[12]

Two sets of neutron powder diffraction intensity data for each $Na_xCoO_2$ sample were collected using the BT-1 high-resolution powder diffractometer at the NIST Center for Neutron Research. For one data set, a Cu (311) monochromator was used to produce a neutron beam of wavelength 1.5403 Å. A Ge(733) monochromator, producing neutrons with wavelength 1.1968 Å was used to collect data in a higher 2θ range to reduce correlations between temperature factors and occupancy parameters. Collimators with horizontal divergences of 15´, 20´, and 7´ of arc were used before and after the monochromator, and after the sample, respectively. The data were collected in two separate banks in steps of 0.05° in the 2θ range 3°-168°. Rietveld refinements of the structures were performed on the two sets of data simultaneously with the GSAS suite of programs.[13] The peak shape was described with a pseudo-Voigt function. The background was fitted with 12 terms in a linear interpolation function. The neutron scattering amplitudes used in the refinements were 0.363, 0.253, and 0.581 ($\times 10^{-12}$ cm) for Na, Co, and O, respectively.

**Results**

Structural models for both $Na_{0.74}CoO_2$ and $Na_{0.38}CoO_2$ have been previously reported.[5, 14] The structures both have crystal symmetry $P6_3/mmm$, differing primarily in the fractional sodium site occupancies. There are two sodium sites, both in trigonal prismatic coordination with oxygen. One prism (Na(2)) shares edges with the neighboring $CoO_6$ octahedra, while the other prism (Na(1)) shares faces with the neighboring $CoO_6$ octahedra. The Na(2) atom is found to be displaced from the center of the trigonal prism in a (2x, x, ¼) site. Previously proposed models have Na(1) on the (0, 0, ¼) site,[5] centered in the trigonal prism, but large thermal parameters for this atom suggested that it might also be displaced from its position.

The structures found here are basically the same as those reported previously but are defined to better precision. The refined hexagonal lattice parameters are found to be $a$=2.8375(1)Å and $c$=10.8781(1)Å for $Na_{0.74}CoO_2$ and $a$=2.8120(1)Å and $c$=11.2289(1)Å for $Na_{0.38}CoO_2$, respectively. The displacements of the Na(2) atoms from the centers of the trigonal prisms are similar to what has been previously found.[5, 15] However the quality of the present data allowed consideration of whether the Na(1) position is similarly displaced. The displacements of the Na(1) atoms from the ideal positions in the center of the prism to (2x, x, ¼) positions was found to be significant in both compounds and yielded conventional magnitude thermal parameters. The displacement is more significant for $Na_{0.38}CoO_2$ with fractional displacements of 17 standard deviations from the ideal site. For $Na_{0.74}CoO_2$, the Na(1) displacements are smaller but are also significant, at 8 standard deviations. The sodium compositions for the two compounds can also be defined to high precision in the refinements, yielding x=0.744(2) and 0.384(2), consistent with expectations based on the known chemistry of these materials. Finally, the quantity and quality of the data allow for simultaneous free refinement of the occupancies and thermal parameters for the oxygen atoms in the close packed layers. The fractional oxygen site occupancies are found to be 1.01(1) and 0.99(2) for $Na_{0.74}CoO_2$ and $Na_{0.38}CoO_2$, respectively. Thus, within the experiment precision, the oxygen sites are found to be filled.

The models presented in Table 1 for the two compositions give excellent fits to the diffraction data for both wavelengths. This can be seen directly in the observed, calculated and difference plots for the refinements, presented in Figures 1 and 2. It can also be seen in the statistical agreement indices presented in Table 1, which are $\chi^2$=1.51 and $\chi^2$=1.38 for all diffraction data for x=0.74 and x=0.38, respectively. Selected bond distances and angles are shown in Table 2: the values are in close agreement with the earlier reported data.[5]

The present work finds both sodium sites, Na(1) and Na(2), to have a distorted trigonal prismatic configuration. Figure 3 presents, as an example, the crystal structure of $Na_{0.74}CoO_2$. The off-center positions result in two types of Na-O bond lengths within the prisms. The displacements of Na(1) and Na(2) in $Na_{0.38}CoO_2$ results in virtually identical bonding environments (left side of Figure 3). The figure also shows the coordination environments in $Na_{0.74}CoO_2$, where the Na(2) site is quite similar in geometry to what is found in x=0.38, but the Na(1) coordination is much less distorted. The fact that the Na(1) site is less displaced in the compound with a higher sodium filling in the plane suggests that ion repulsion within the layer forces it to become on-center: the implication is that the naturally occurring positions within the prism, without the influence of in-plane Na-Na repulsion, are off-center. The fractional occupancies of the oxygen sites indicate that there is no oxygen deficiency at an uncertainty of 1 to 2%. This result demonstrates that the sodium content is an accurate measure of the formal oxidation state of Co in $Na_xCoO_2$. That is +3.26 and +3.62 for $Na_{0.74}CoO_2$ and $Na_{0.38}CoO_2$, respectively. As a result, the level of electron doping is well defined by the sodium content. The possibility of oxygen loss during the Na deintercalation process is excluded by the present results on $Na_{0.38}CoO_2$. The only prospect for oxygen deficiency in the superconducting phase would, therefore, have to be oxygen vacancies formaing during the hydration process. This, in fact, has been previously reported.[8]

**Conclusions**

High precision powder diffraction analysis of the structures of $Na_{0.75}CoO_2$ and $Na_{0.38}CoO_2$ indicates that there is no oxygen non-stoichiometry either in the thermodynamically obtained composition $Na_{0.75}CoO_2$ or in the deintercalated product $Na_{0.38}CoO_2$. A previous high resolution diffraction study of $Na_{0.5}CoO_2$ also found no oxygen deficiency. It can therefore be concluded that in the range of compositions that display the interesting electronic and magnetic properties in the $Na_xCoO_2$ system the formal charge on Co is given only by the sodium content. Conclusions about the presence or absence of $H^+$, $H_3O^+$ or oxygen vacancies in the superconducting superhydrate cannot be drawn from the current work. However, if oxygen vacancies are indeed present in the superconducting material the fact that they are not present in the $Na_{0.38}CoO_2$ host material implies they would have to be formed during the hydration process.[8] The refinements of the two structures also revealed that the two sodium sites have distorted trigonal prismatic configurations. The finding that off-center positions of the Na are preferred in this system for both types of Na sites, those that share edges with the $CoO_6$ octahedra and those that share faces, suggests that these are energetically preferred positions for Na within the layer. Why this is the case is an interesting crystal-chemical question that remains to be answered.

**Acknowledgements**

This research was supported by the US Department of Energy Division of Basic Energy Sciences.

**Table 1.** Structure parameters of $Na_{0.74}CoO_2$ and $Na_{0.38}CoO_2$.

| compound | atom | site | *x* | *y* | *z* | $U_{iso}$x100 | *Occ* |
|---|---|---|---|---|---|---|---|
| [a]$Na_{0.74}CoO_2$ | Co | *2a* | 0 | 0 | 0.5 | 0.54(5) | 1 |
| | Na(1) | *6h* | 0.051(8) | 0.025(4) | 0.25 | 1.5(1) | 0.071(2) |
| | Na(2) | *6h* | 0.559 (2) | 0.2796(8) | 0.25 | 1.5(1) | 0.177(2) |
| | O | *4f* | 1/3 | 2/3 | 0.09047(7) | 0.76(1) | 1.01(1) |
| [b]$Na_{0.38}CoO_2$ | Co | *2a* | 0 | 0 | 0.5 | 0.72(5) | 1 |
| | Na(1) | *6h* | 0.104(6) | 0.052(3) | 0.25 | 1.5(2) | 0.038(2) |
| | Na(2) | *6h* | 0.571(3) | 0.285(1) | 0.25 | 1.5(2) | 0.090(2) |
| | O | *4f* | 1/3 | 2/3 | 0.08585(5) | 0.69(2) | 0.99(2) |

[a]$\chi^2$=1.51; for $\lambda$=1.5403Å: wRp=7.59%; Rp=6.34%; Re=5.02%; for $\lambda$=1.1968Å: wRp=6.46%; Rp=5.28%, Re=4.28%

[b]$\chi^2$=1.38; for $\lambda$=1.5403Å: wRp=5.95%; Rp=4.99%; Re=4.31%; for $\lambda$=1.1968Å: wRp=5.52%; Rp=4.51%; Re=4%;

**Table 2.** Selected bond distances and angles for $Na_{0.74}CoO_2$ and $Na_{0.38}CoO_2$.

| Bond distance (Å) / Angle (degrees) | $Na_{0.74}CoO_2$ | $Na_{0.38}CoO_2$ |
|---|---|---|
| Co-O × 6 | 1.9111(4) | 1.8882(3) |
| Na(1)-O × 4 | 2.347(6) | 2.389(2) |
| × 2 | 2.47(1) | 2.616(5) |
| Na(2)-O × 4 | 2.309(1) | 2.384(4) |
| × 2 | 2.575(3) | 2.63(1) |
| O-Co-O | 84.13(3) | 83.74(2) |

**Figures Captions**

**Figure 1.** Observed (crosses) and calculated (solid line) neutron diffraction intensities for $Na_{0.74}CoO_2$. Vertical bars show the Bragg peak positions. The data for $\lambda$=1.5403Å are shown in the inset.

**Figure 2**. Observed (crosses) and calculated (solid line) neutron diffraction intensities for $Na_{0.38}CoO_2$. Vertical bars show the Bragg peak positions. The data for $\lambda$=1.5403Å are shown in the inset.

**Figure 3.** The crystal structure of $Na_{0.74}CoO_2$. Edge-shared $CoO_6$ octahedra are separated by layers of sodium ions. The two sodium sites in this structure are distorted triangular pyramids, represented on the right.

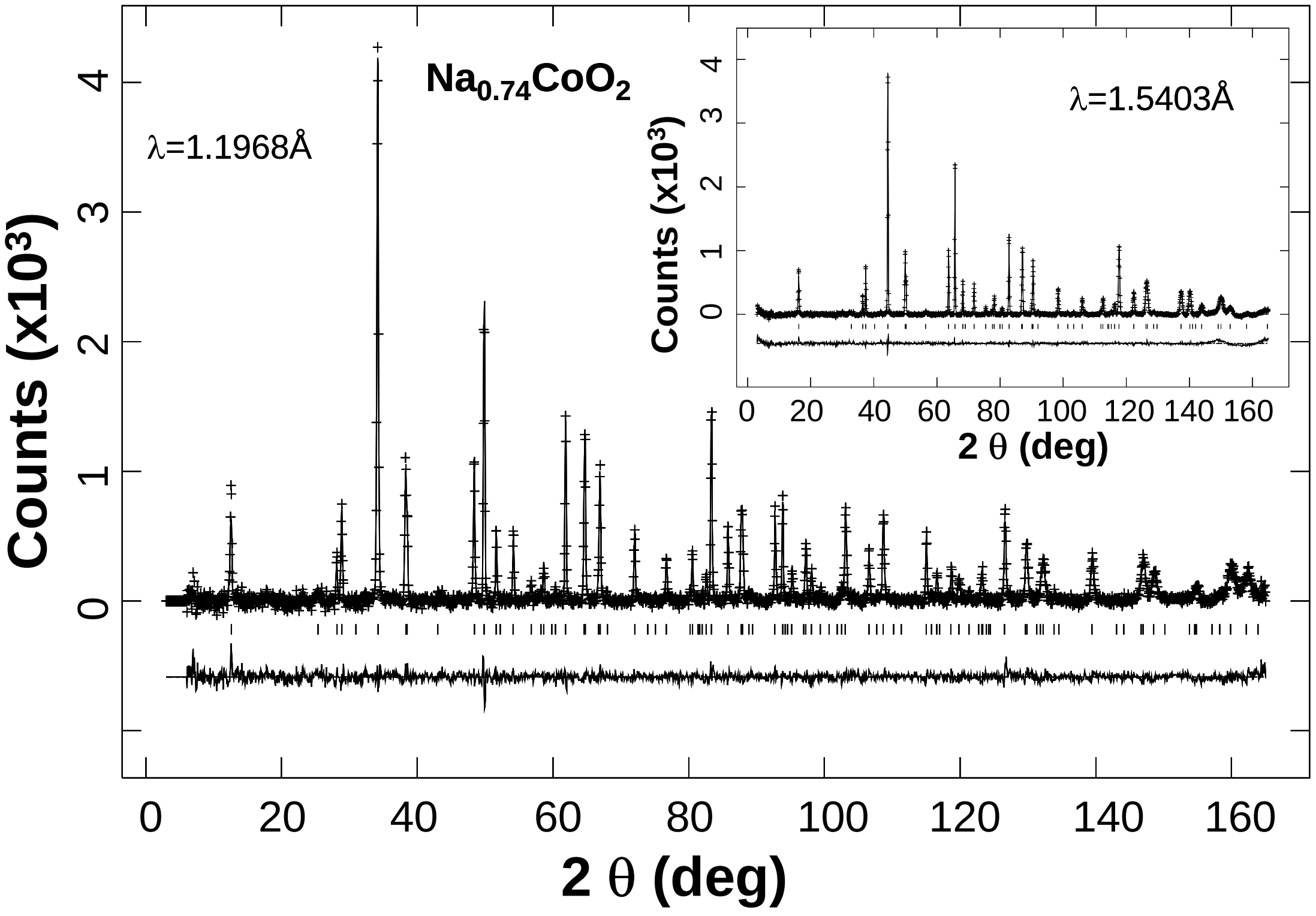

**Figure 1**

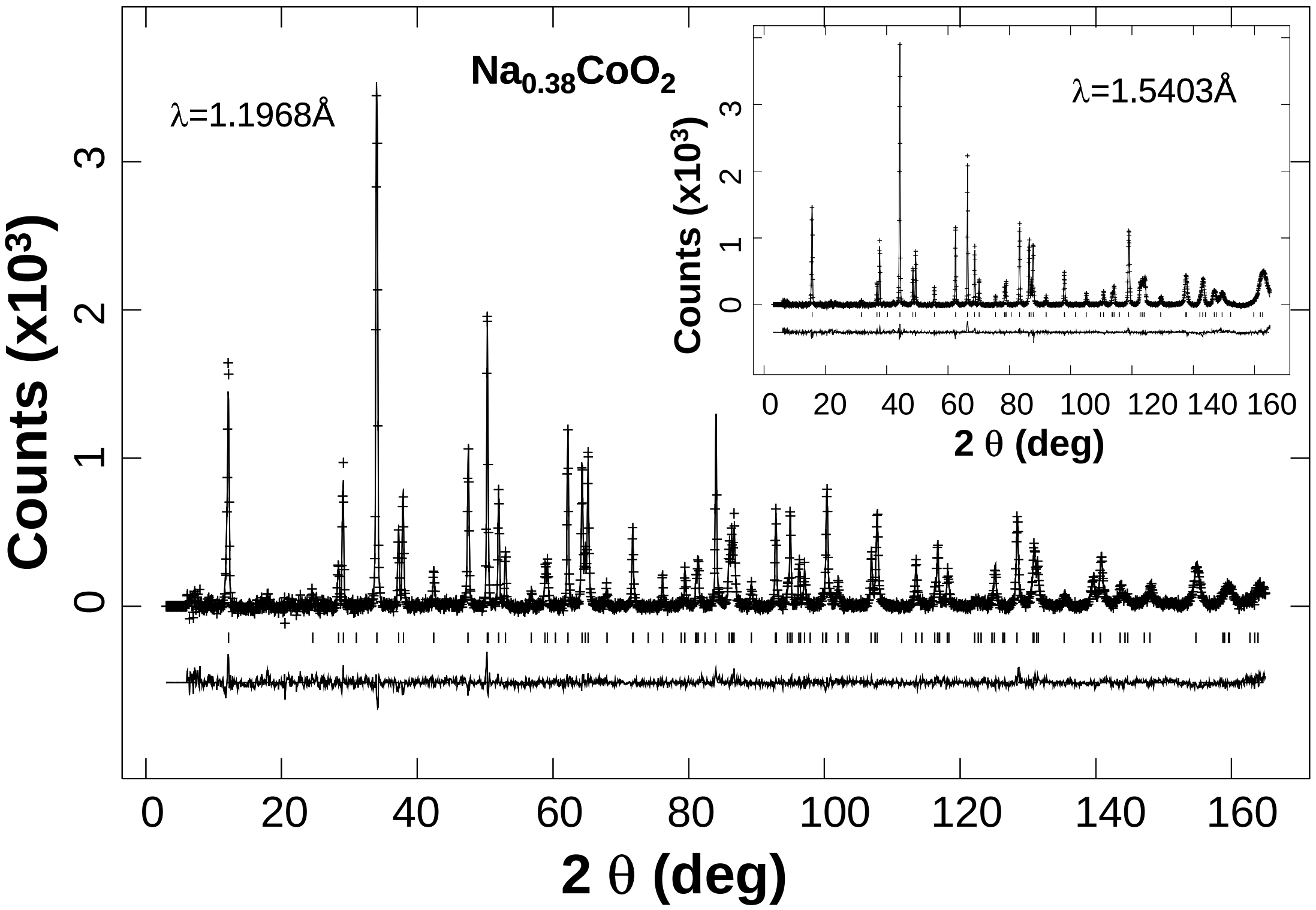

**Figure 2**

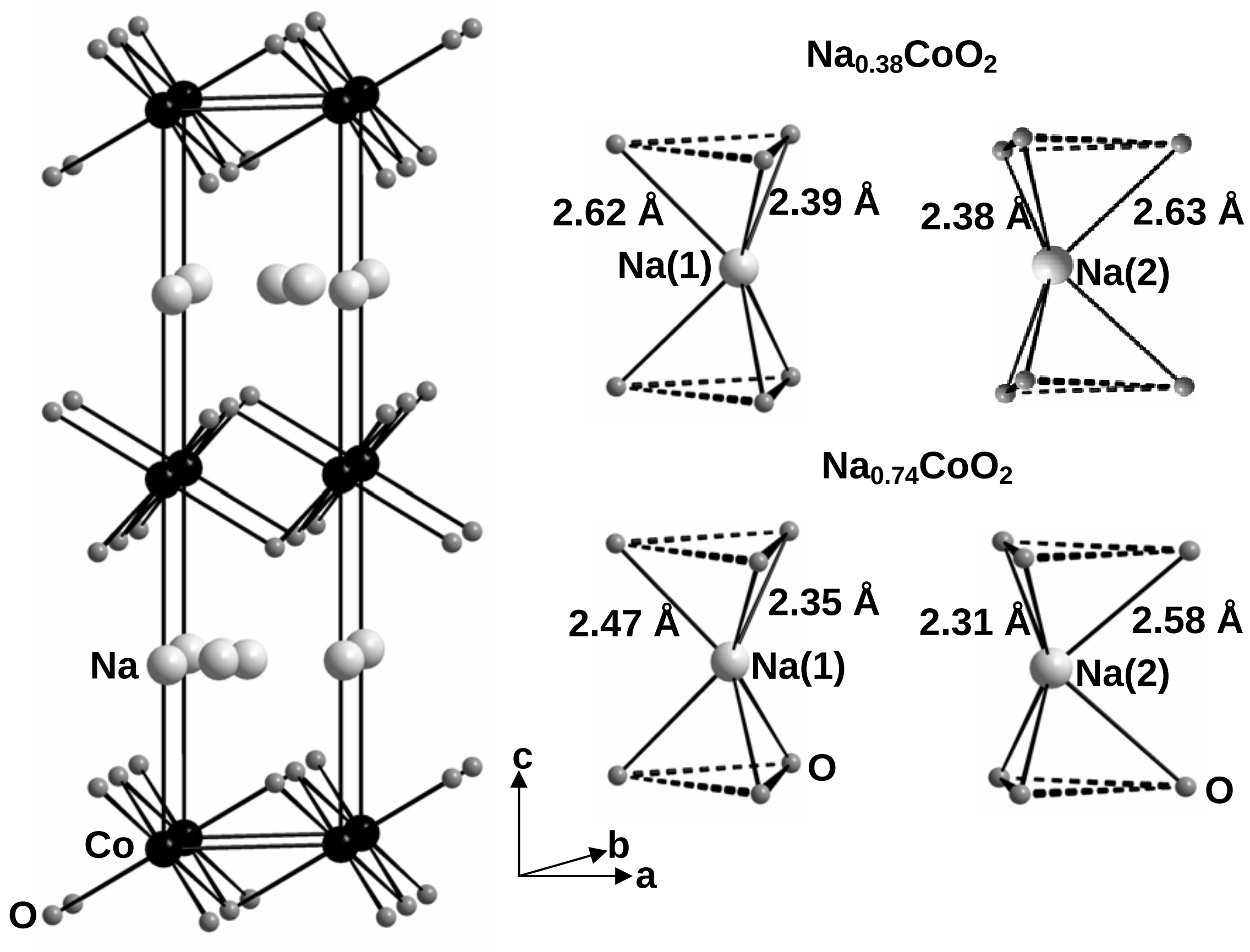

**Figure 3**